\documentclass[twocolumn,tighten]{aastex631}

\usepackage{amsmath,amstext}
\usepackage{mathrsfs}
\usepackage{float}

\usepackage{afterpage}
\usepackage{xcolor}
\usepackage{threeparttable}

\usepackage{graphicx}
\usepackage{epsfig}

\usepackage{booktabs}
\usepackage{multirow}

\usepackage{dcolumn}% Align table columns on decimal point
\usepackage{bm}% bold math
\usepackage{accents}

\begin{document}

\title{A new covariant formalism for kinetic plasma simulations in curved spacetimes}

\author{Tyler Trent}
\affiliation{School of Physics, Georgia Institute of Technology, 837 State St NW, Atlanta, GA 30332, USA}
\affiliation{Departments of Astronomy and Physics, University of Arizona, 933 N. Cherry Ave., Tucson, AZ 85721, USA}
%\correspondingauthor{Trent}
\email{ttrent@arizona.edu}

\author{Pierre Christian}
\affiliation{Physics Department, Fairfield University, 1073 N Benson Rd, Fairfield, CT 06824}

\author{Chi-kwan Chan}
\affiliation{Departments of Astronomy and Physics, University of Arizona, 933 N. Cherry Ave., Tucson, AZ 85721, USA} 
\affiliation{Data Science Institute and Program in Applied Mathematics, University of Arizona, 1230 N. Cherry Ave., Tucson, AZ 85721}

\author{Dimitrios Psaltis}
\affiliation{School of Physics, Georgia Institute of Technology, 837 State St NW, Atlanta, GA 30332, USA}

\author{Feryal \"Ozel}
\affiliation{School of Physics, Georgia Institute of Technology, 837 State St NW, Atlanta, GA 30332, USA}

\begin{abstract}
Low density plasmas are characterized by a large scale separation between the gyromotion of particles around local magnetic fields and the macroscopic scales of the system, often making global kinetic simulations computationally intractable. The guiding center formalism has been proposed as a powerful tool to bridge the gap between these scales. Despite its usefulness, the guiding center approach has been formulated successfully only in flat spacetimes, limiting its applicability in astrophysical settings. Here, we present a new covariant formalism that leads to kinetic equations in the guiding center limit that are valid in arbitrary spacetimes. Through a variety of experiments, we demonstrate that our equations capture all known gyro-center drifts while overcoming one severe limitation imposed on numerical algorithms by the fast timescales of the particle gyromotion. This formalism will enable explorations of a variety of global plasma kinetic phenomena in the curved spacetimes around black holes and neutron stars. 
\end{abstract}

\section{Introduction}
Rarefied plasmas are ubiquitous in a diverse range of astrophysical systems, from the heliosphere to the intracluster medium and the accretion flows around black holes. Magnetohydrodynamics (MHD), sometimes in its general relativistic formulation (GRMHD), is often the method of choice when modeling these systems. While useful for understanding their overall dynamics, MHD makes a number of assumptions and approximations that are often not valid in the low density regime. For example, fluid approaches cannot capture phenomena that arise from large mean-free-paths ($\lambda_{\rm mfp}$) of particles, pressure anisotropies, and non-ideal acceleration and dissipation effects. All of these effects determine the observational appearance of plasmas and impact the interpretation of astrophysical observations. Of recent interest are imaging observations of nearby black holes with the Event Horizon Telescope~\citep{EHT_M87_1,EHT_SGRA_1} and the electromagnetic counterparts of neutron-star mergers~\citep{LIGO_NS}. 

To resolve the shortcomings of the fluid models, one needs to rely on kinetic approaches and solve for the individual motions of charged particles and the fields they produce. While able to resolve much of the microphysics that fluid simulations are unable to, such approaches have been limited to local simulations that study regions of an astrophysical system (for plasmas around compact objects, see, e.g., \citealt{Ball_2018_july, Hakobyan_2019, Ball_2019}) or to global simulations that use nonphysical parameters \citep{Parfrey_2019, Crinquand_2022, Galishnikova_2023}. One of the critical limitations is the large scale separation between the microscopic radius of the charged particle gyro-motion $\rho$ and the macroscopic size of the system $R$, which severely constrains the dynamical range that can be numerically simulated. However, this same scale separation also provides an opportunity to reformulate the problem in a computationally tractable way using a guiding center approach. 

The guiding center formalism decomposes the motion of a charged particle into a fast gyration around a guiding center (sometimes referred to as gyro-center) and the slower motion of the guiding center itself (sometimes referred to as the drift motion). This approach assumes that the electromagnetic field is slowly varying in space when compared to the gyro-radius and slowly varying in time when compared to the gyro-period. By solving for only the guiding center of the particle, kinetic models can use timesteps larger than the gyro-period and are thus able to solve for the guiding center motion over larger scales, overcoming the challenge of scale separation. 

\begin{figure}
    \centering
    \includegraphics[width=0.48\textwidth]{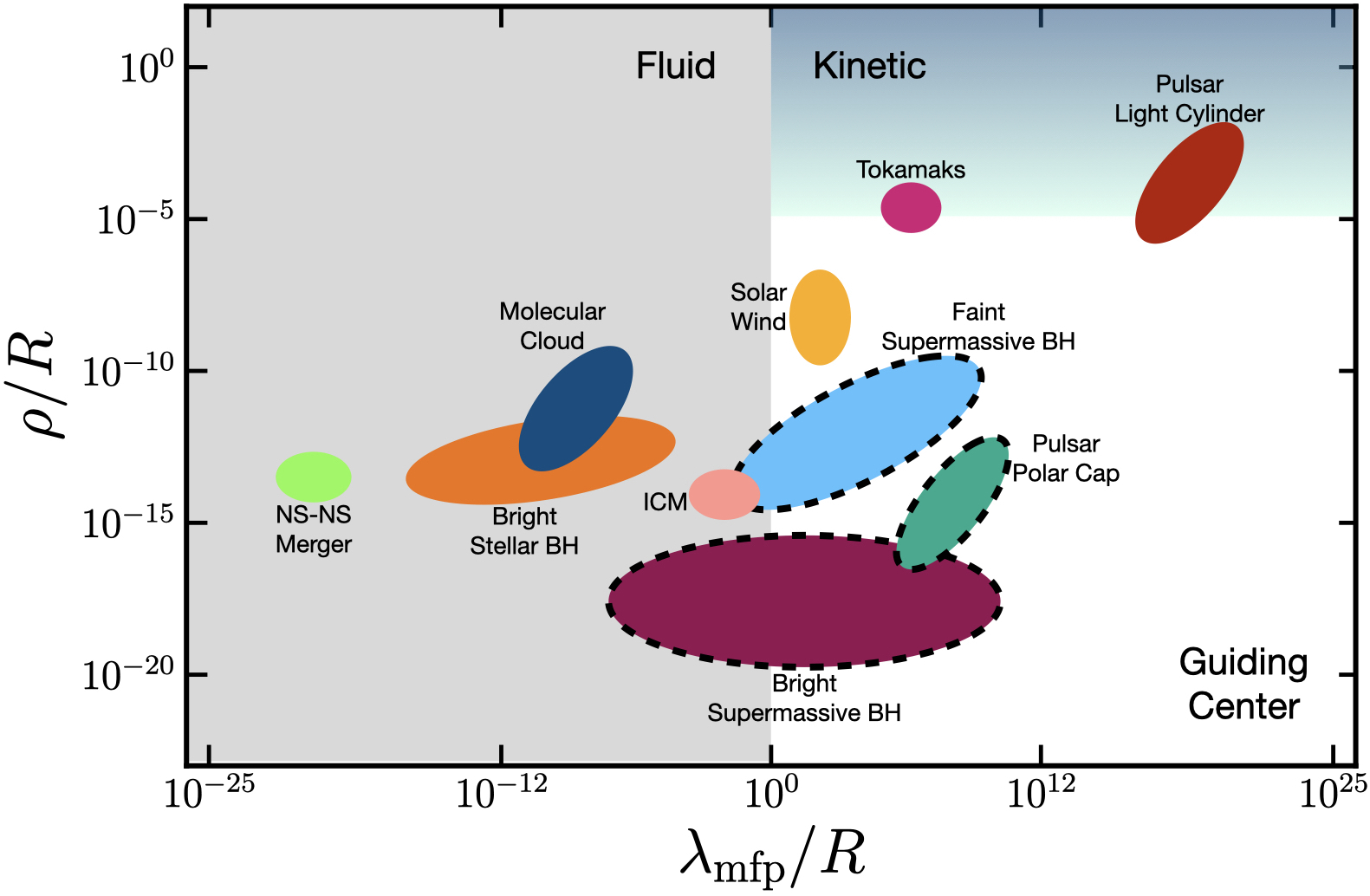}
    \caption{The physical conditions encountered in plasmas in a variety of astrophysical settings; $\rho$ denotes the radius of gyromotions, $\lambda_{\rm mfp}$ the collisional mean-free path of particles, and $R$ the macroscopic scale of each system. A fluid approach is applicable for modeling systems in the gray shaded region. The conditions in the blue shaded region necessitate a kinetic approach. The guiding center formalism developed here enables accurate integrations of charged particle trajectories in all the systems in the white region. Dashes outline systems for which general relativistic effects require a covariant formulation.}
    \label{fig:advertisement}
\end{figure}

Figure~\ref{fig:advertisement} shows the physical conditions present in a variety of astrophysical, solar, and terrestrial plasmas. It also identifies regions for which the large mean free paths ($\lambda_{\rm mfp}/R\gtrsim 1)$ necessitate a kinetic approach but the scale separation ($\rho/R\ll 1$) allows the application of the guiding center approximation. Adopting the latter would enable some investigations that have not been possible to date. 

In flat spacetimes, there exists a widely used set of guiding center equations \citep{Northrop_book} that have been successfully implemented in studies, often in conjunction with a background MHD model~(see, e.g, \citealt{Gordovskyy_2010, Threlfall_2016, Ripperda_2017, Gordovskyy_2020} for studies of solar and astrophysical flares). Despite the usefulness and the demonstrated accuracy of the guiding center approach in this regime, there has not been a successful covariant formalism to date, which is necessary for applications in relativistic and compact astrophysical systems. In the traditional formalism, one uses approximate methods to integrate analytically the acceleration equation for the particle once over a gyro-period and obtain an equation for the drift velocity of the guiding center, which can then be solved numerically. This expression typically contains terms related to the curvature and gradient of the $B$ field, as well as the electric, $\vec{E} \times \vec{B}$, and gravitational, $\vec{g} \times \vec{B}$, drifts. However, the step of integrating analytically the acceleration equation cannot be generalized in the same manner to a covariant form because the geodesic equation has nonlinear terms in velocity. An alternate approach of solving the problem in the local Lorentz frame (as done, e.g., in \citealt{Bacchini_2020}) also fails because terms that involve the gradients of the electromagnetic field tensor and the gravitational drift cannot be written in terms of quantities evaluated in the local Lorentz frame~\footnote{One attempt by \citet{Beklemishev_2004} leaves the equations in a closed form, but one which is computationally intractable. In \citet{Bacchini_2020}, the guiding center equations were solved in curved spacetimes, but only the $\vec{E}\times\vec{B}$ drift was incorporated, in a manner that cannot be generalized to a fully covariant description that includes curvature and gravitational drifts.}.

In this letter, we derive a new set of fully covariant guiding-center equations of motion that incorporate all of the drift mechanisms captured in the standard guiding center set of equations \citep{Northrop_book}, including the gravitational drift. The key insight in this approach comes from the realization that, although it is not possible to obtain a covariant equation for the drift velocity of the guiding center, one can derive an equation for its acceleration while still integrating out the gyromotion. The resulting second order differential equation can then be solved numerically without significant increase in the computational cost. 

\begin{figure}
    \centering
    \includegraphics[width=0.45\textwidth]{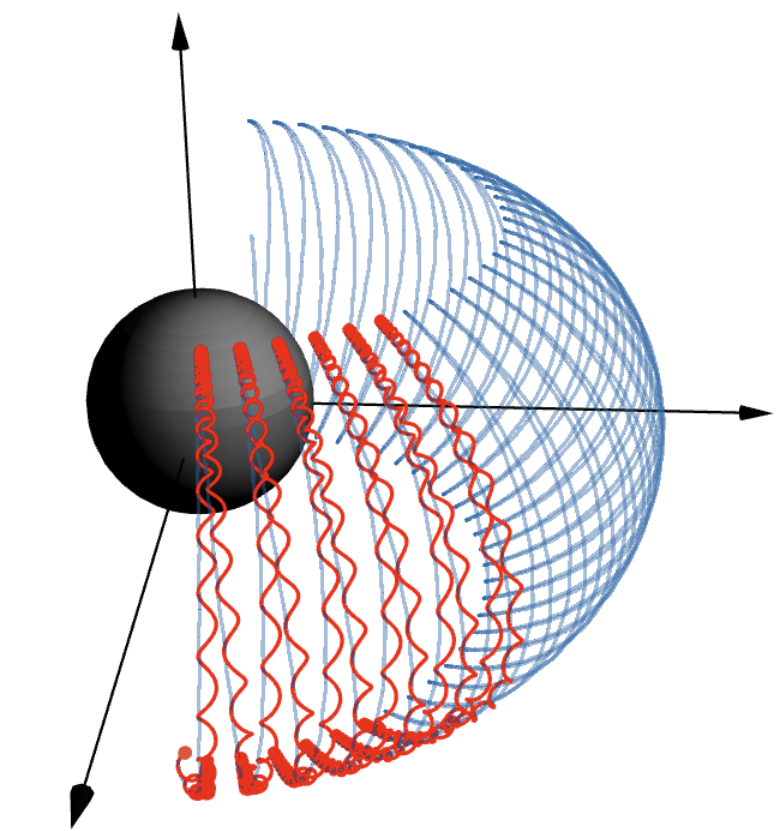}
    \caption{The trajectory of a charged particle moving in a magnetic dipole in a Schwarzschild spacetime, calculated by solving the full equations of motion (red line) as well as the equations of the guiding center derived here (blue line). The particle experiences mirroring between the two magnetic poles and an azimuthal drift. Even for the exaggerated conditions shown here, where the gyration is clearly visible in the trajectory of the particle, the guiding center equations describe accurately the drift of the particle motion.}
    \label{fig:traj}
\end{figure}

Figure~\ref{fig:traj} illustrates with an example the guiding center approach. A charged particle travels in a dipole magnetic field around a compact star, both mirroring between the two magnetic poles and drifting in the azimuthal direction. The figure shows both the full gyrating motion of the particle (in red) as well as the motion of its guiding center that results from the approach we derive in this Letter. For visual purposes, the parameters were chosen such that gyrating motion is visible in the full motion of the particle, which is highly exaggerated compared to realistic systems (see Fig~1). Even in this case, the guiding center approach shows a remarkable agreement with the full solution.

\smallskip

\section{The Covariant Guiding Center Equation of Motion}
The covariant equation of motion for a charged particle in a general spacetime with an arbitrary electromagnetic field is given by
\begin{equation}
\label{eqn:full EOM}
    \frac{d^2x^\alpha}{d\tau^2} = -\Gamma^\alpha_{\mu\nu} \frac{dx^\mu}{d\tau} \frac{dx^\nu}{d\tau} +\frac{q}{m}F^{\alpha}_{~ \beta}\frac{d x^\beta}{d\tau},
\end{equation}
where $x^\alpha$ is the position of the particle, $\tau$ is the proper time, $\Gamma^\alpha_{\mu\nu}$ is the Christoffel symbol, $q/m$ is the charge-to-mass ratio of the particle, and $F^{\alpha}_{~\beta}$ is the electromagnetic field tensor. The goal is  to decompose the motion of the particle into a fast gyromotion and a slow drift of the guiding center and integrate out the gyromotion analytically. 

Expanding the electromagnetic field tensor around the position $\chi^\mu$ of the guiding center, we obtain
\begin{eqnarray}
\label{eqn:expandedEOM}
    \frac{d^2x^\alpha}{d\tau^2} &=& \left.-\Gamma^\alpha_{\mu\nu}\right\vert_{\chi} \frac{dx^\mu}{d\tau} \frac{dx^\nu}{d\tau} +\frac{q}{m}\left.F^{\alpha}_{~ \beta}\right\vert_{\chi}\frac{d x^\beta}{d\tau}\nonumber\\
    &&\qquad
    +\frac{q}{m}\left.\frac{\partial F^{\alpha}_{~ \beta}}{\partial x^\mu}\right\vert_{\chi}\left(x^\mu-\chi^\mu\right)\frac{d x^\beta}{d\tau}. 
\end{eqnarray}
The first and the third term in this equation are first order in $\rho/R$ and can be neglected to zeroth order when the following three conditions are satisfied (see also \citealt{Vandervoort_1960}): \\
\noindent 1. The gyroradius $\rho$ is significantly smaller than the characteristic scale over which the electromagnetic field varies
\begin{equation}
    \rho\ll|F^\alpha_{~\beta}| / \left|\frac{\partial F^\alpha_{~\beta}}{\partial x^\mu}\right|\;.
\end{equation}
\noindent 2. The particle can drift for many gyroperiods\\ before the field changes considerably
\begin{equation}
    \frac{1}{\omega}\left|\frac{\partial\chi^\nu}{\partial \tau}\right|\ll|F^\alpha_{~\beta}| / \left|\frac{\partial F^\alpha_{~\beta}}{\partial x^\mu}\right|\;.
\end{equation}
\noindent 3. The effect of the spacetime curvature on the motion of the particle is weaker than that of the electromagnetic field
\begin{equation}    \left|\Gamma^\alpha_{\mu\nu}\frac{dx^\mu}{d\tau} \frac{dx^\nu}{d\tau}\right|\ll \frac{q}{m} \left|F^\alpha_{~\beta}\frac{dx^\beta}{d\tau}\right|\;.
\end{equation}
In the above expressions, the symbol $|~|$ denotes the magnitude of a typical component.

To zeroth order, the reduced Eq.~(\ref{eqn:expandedEOM}) becomes a homogeneous differential equation with constant coefficients; we can write its full solution in terms of the eigenvectors and eigenvalues of the $F^\alpha_{~\beta}$ tensor \citep{Vandervoort_1960, Fradkin_1978}. There are two imaginary and two real eigenvalues, which we denote by $\{i\omega,-i\omega,\lambda,-\lambda\}$, and their corresponding eigenvectors as $\{\sigma,\delta,\psi,\Upsilon\}$. The eigenvalues can be conveniently written in terms of the field tensor invariants and the eigenvectors can then be solved using the Cayley-Hamilton theorem \citep{Fradkin_1978}. The two imaginary eigenvalues correspond to the gyromotion of the particle with angular frequency $\omega$, as expected, while the two real eigenvalues describe the drift of the guiding center. 

In this limit, the full solution of the particle motion can be expressed as a linear combination of the four eigenvectors. Because of the comparatively small scale of the gyroradius, to first order, the presence of a curved spacetime and a spatially varying electromagnetic field will influence the motion of the guiding center but not the gyration. To obtain the solution in this more general case, we, therefore, only use the eigenvectors that correspond to the gyromotion but do not prescribe the motion of the guiding center in terms of the other two eigenvectors. Instead, we write 
\begin{equation}
\label{eqn:four-position definition}
    x^\alpha(\tau) = \rho_0\sqrt{\frac{\omega_0}{\omega}}e^{i\omega\tau}\,\sigma^\alpha
    +\rho^*_0\sqrt{\frac{\omega_0}{\omega}}e^{-i\omega\tau}\,\delta^\alpha
    +\chi^\alpha\;,
\end{equation}
where $\rho_0\equiv -(i\delta_\beta/\omega)(dx^\beta/d\tau)\vert_{\tau=0}$ is the gyro-radius; in this expression, $\rho_0$ may be complex and thus does not correspond directly to the usual definition.

In order to obtain an equation for the guiding center position, we insert our ansatz (Eq.~[\ref{eqn:four-position definition}]) into the equation of motion given in Eq.~(\ref{eqn:expandedEOM}) and expand all terms to first order in $\rho/R$. Finally, we time average the differential equation over one gyro-period, zeroing out all terms that are oscillatory. The resulting equation of motion for the guiding center becomes\footnote{Equation~\eqref{eqn:guiding center EOM} can be written in a manifestly covariant form by combining the second and fourth term into $i\omega_0\rho^2_0 \frac{q}{m}\nabla_\mu F^\alpha_{~\beta}(\sigma^\beta\delta^\mu-\sigma^\mu\delta^\beta)$.}
\begin{align}
\label{eqn:guiding center EOM}
    \frac{d^2\chi^\alpha}{d\tau^2} =
    &-\Gamma^\alpha_{\mu\nu} \biggl( \frac{d\chi^\mu}{d\tau}\frac{d\chi^\nu}{d\tau}
    +2\omega\omega_0\rho_0^2\sigma^\mu \delta^\nu\biggr)\\ \nonumber
    &+\frac{q}{m}F^\alpha_{~\beta}\frac{d\chi^\beta}{d\tau}
    + i\omega_0\rho^2_0 
    \frac{q}{m}\frac{\partial F^\alpha_{~\beta}}{\partial x^\mu}(\sigma^\beta\delta^\mu-\sigma^\mu\delta^\beta)   
\end{align}
The acceleration terms in Eq.~\eqref{eqn:guiding center EOM} correspond to each of the familiar drift mechanisms. The first two terms containing the Christoffel symbols are responsible for the gravitational drift. The last term is responsible for drift due to a non-constant electromagnetic field which includes the $\nabla B$ drift. Note that, because the eigenvectors are complex, the last term in Eq.~\eqref{eqn:guiding center EOM} is indeed real. 

\section{Application and Verification}
In order to demonstrate that the new covariant, guiding-center equations we derived above account for all known drift mechanisms, we devised a set of test problems in various configurations. Specifically, we consider the constant electromagnetic field in flat spacetime, which results in an $\vec{E} \times \vec{B}$ drift, the dipole magnetic field in a flat spacetime, which yields a $\nabla B$ drift, and a dipole magnetic field in curved spacetime, which also contains a gravitational drift. Because our goal here is to demonstrate the applicability of the guiding center equation and not to explore and optimize a numerical particle pusher for solving it, we use a simple fourth order Runge-Kutta integrator to solve eq.~(\ref{eqn:guiding center EOM}). We then compare the result to the solution of the full equation of motion~(\ref{eqn:full EOM}) for the charged particle. We obtain the latter either analytically or numerically. Hereafter, we set $G=c=1$ and absorb the value of $q/m$ into the magnitude of the magnetic field. 

\section{Constant Electromagnetic Field in Flat Spacetime}
We first study the motion of the guiding center of a charged particle in a constant electromagnetic field in a flat spacetime and compare it to the analytic solution. For this configuration, the guiding center is expected to drift with a velocity  $v_E = (\vec{E}\times\vec{B})/ B^2$. We initialize the particle at an arbitrary position $x^\alpha=(0,5\sqrt{2},5\sqrt{2},0)$ and with a velocity $u^\alpha=(u^t,0,u^y,2)$. We set the components of the electric and magnetic field to $E^i=(1,0,-0.05)$ and $B^i=(0,0,B_0)$, respectively. By varying the $u^y$ component of the velocity and the $B^z$ component of the magnetic field, we explore different sizes of the gyroradius and, therefore, test different regimes in scale separation. In each case, we infer the $u^t$ component of the velocity from the requirement $u^a u_a=-1$.

\begin{figure}
    \centering
    \includegraphics[width=0.46\textwidth]{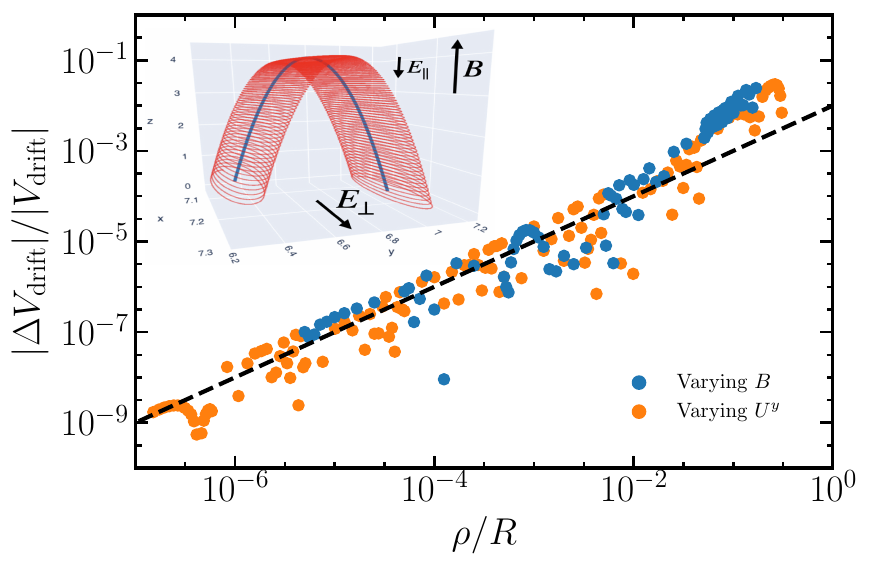}
    \caption{Fractional difference between the average drift velocity calculated numerically in the guiding center limit and the analytic drift velocity for a charged particle in a constant electromagnetic field, as a function of the gyroradius $\rho$ divided by the macroscopic scale $R$ of the system, in a flat spacetime (see insert for the setup). In this configuration, the only drift experienced by the particle is proportional to $\vec{E}\times\vec{B}$. Different points show different magnetic field strengths (blue) and particle velocities (orange), both of which alter the gyroradius. The truncation error in the drift velocity introduced by the guiding-center equations depends linearly on gyroradius, as expected.}
    \label{fig:flat_cnst_err}
\end{figure}

With this initial velocity and under the influence of the parallel and perpendicular components of the electric field (see insert in Figure~\ref{fig:flat_cnst_err}), the particle goes through an arc like motion before re-crossing the $z=0$ plane. The horizontal displacement on this plane is due to the $E \times B$ drift, which we estimate by dividing the guiding center displacement by the amount of elapsed coordinate time. 

We show in Fig.~\ref{fig:flat_cnst_err} the fractional difference between the average drift velocity measured from the guiding center calculation and the analytic drift velocity, as a function of the gyroradius. As expected, the guiding center equations become increasingly more accurate with decreasing gyroradius. Because we kept terms to first order in gyroradius in the guiding center equations, we expect the truncation error to be of second order. However, the leading order term in the drift velocity is first order in gyroradius and, therefore, the truncation error in drift velocity is only one order higher. This is why Fig.~\ref{fig:flat_cnst_err} shows a linear dependence of the fractional error on gyroradius. The figure also shows that, as expected, the degree of approximation of the guiding center solution depends only on the magnitude of the gyroradius and not on the magnitude of the magnetic field or of the particle velocity. 

\begin{figure}
    \centering
    \includegraphics[width=0.45\textwidth]{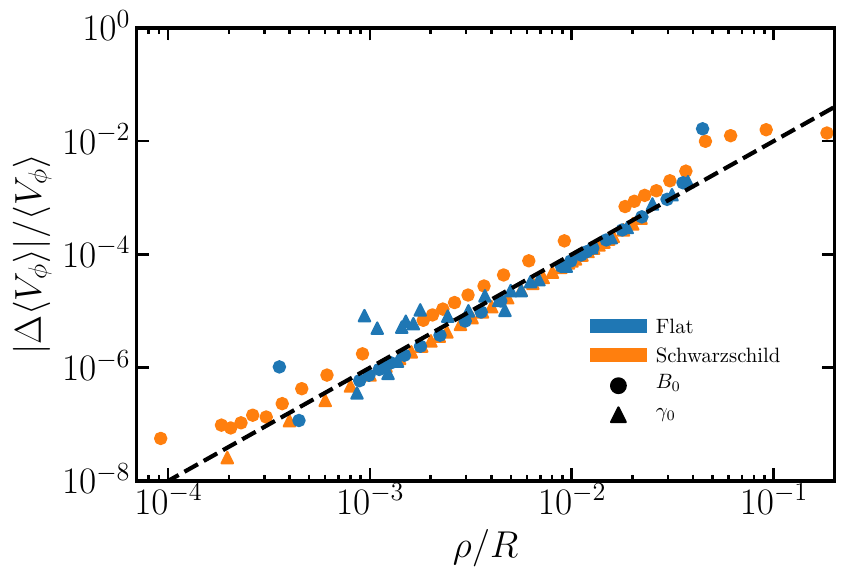}
    \caption{Fractional difference between the average azimuthal drift velocity calculated in the guiding center limit and using the full equation of motion of a charge in a magnetic dipole in flat and Schwarzschild spacetimes (see Fig.~\ref{fig:traj} for the setup). Other details as in Fig.~\ref{fig:flat_cnst_err}. In this configuration, the truncation error depends quadratically on gyroradius.
     }
    \label{fig:magdip_err}
\end{figure}

\section{Dipole Magnetic Field in Flat and Schwarzschild Spacetimes}
In this application, we consider the motion of a charged particle in a dipole magnetic field, both in a flat and in a Schwarzschild spacetime. In both settings the particle experiences magnetic mirroring between the two magnetic poles as well as azimuthal drift, due to the gradient in the magnetic field strength. This is the configuration shown earlier in Fig.~\ref{fig:traj}. In the Schwarzschild case, the particle also experiences gravitational drift. 

We implement the magnetic dipole in terms of its vector potential \citep{Bacchini_2019, Takahashi_2009} 
\begin{align}
\label{eqn:schwarzschild dipole}
    A_\phi = \frac{3}{4}B_0\sin^2(\theta) \Bigg[2(r+1)-r^2\log\biggl(\frac{r}{r-2}\biggr)\Bigg]\;,
\end{align}
where all radii are written in terms of $GM/c^2$ and $M$ is the mass of the central object. For the flat spacetime configuration, we take the limit $r\gg 1$.

We choose an initial position of the particle on the equatorial plane at a radius $r=10 M$ and with an initial velocity at a pitch angle $\Theta=\pi/4$, i.e., we set $u^\alpha=[(-g_{tt})^{-1/2}\gamma_0,(-g_{tt})^{1/2}\beta\gamma_0\sin\Theta,\beta\gamma_0 r^{-1}\sin\Theta,0]$, where $\beta = (1-1/\gamma_0^2)^{1/2}$
and $g_{tt}$ is the $tt-$component of the metric. The magnitude $B_0$ of the magnetic field and the Lorentz factor $\gamma_0$ determine, as before, the particle gyroradius. 

In this configuration, there is no analytic solution for the drift velocity of the particle. In order to test the covariant guiding center equations and explore their convergence in $\rho/R$, we compare our results to those obtained from integrating the full particle trajectory. In particular, we estimate numerically the drift velocity in the azimuthal direction by tracking the times and azimuths of the successive equatorial crossings of the particle in both solutions. In Fig.~\ref{fig:magdip_err}, we show the fractional difference between the two estimates of the azimuthal drift velocity as a function of the gyroradius. Even in this configuration that incorporates all known drift mechanism, the guiding center equations maintain high accuracy, with a truncation error that scales as $(\rho/R)^2$. This steeper dependence compared to the previous configuration  likely originates from a hidden symmetry in the problem.  

We emphasize here that the results shown in Figs.~\ref{fig:flat_cnst_err}-\ref{fig:magdip_err} aim to demonstrate the correct limiting behavior of our covariant guiding-center equations and are not convergence plots of a numerical solver. (The latter would have shown the difference between two solutions as a function of numerical resolution and not of the expansion parameter, $\rho/R$, in the equations).

\section{Conclusions}

In this paper, we developed a new covariant guiding center formalism for the motion of charged particles in general spacetimes that account for electric, magnetic, and gravitational drifts. We showed that the solution to these equations match those of the full equations of motion at the limit $\rho/R \rightarrow 0.$ Our approach allows integrating particle trajectories with time steps that are set by the macroscopic length scales $R$ and not by the gyroradius $\rho$. This leads to an increase of order $R/\rho$ in the computational efficiency when numerically solving for particle trajectories in arbitrary spacetimes and electromagnetic field configurations. 

The new guiding center formalism will allow us to explore a number of interesting plasma phenomena in astrophysics affected by the long mean-free paths of charges and the large scale separation between the system size and the gyroradii of charged particles. In particular, phenomena in settings where a background magnetic field is determined by currents elsewhere in the system can be fruitfully simulated with the guiding center formalism. Another application is the motion of charges that contribute negligibly to the overall dynamics and the generation of the electromagnetic field in the flow, but are important for determining the radiative and observational signatures of the systems. 

The trapping of non-thermal particles in accretion flows is one such application. Non-thermal electrons in accretion flows are thought to originate from magnetic reconnection and can potentially explain the bright flares observed from low-luminosity systems, such as Sgr~A*. Recent observations reveal that, despite their long mean-free paths, these non-thermal particles are trapped in quasi-coherent compact structures that appear to orbit around the black hole \citep{GRAVITY}. Calculating the spatial distribution of non-thermal particles and understanding the mechanism that confines the flaring emission to a compact region necessitates a kinetic approach that will follow the trajectories of the individual non-thermal particles in the background GRMHD flow. The new guiding center formalism can provide an optimal tool for simulating and understanding such systems. 

\begin{acknowledgements}
    We thank Gabriele Bozzola, Dirk Heumann, and Matthew Golden for useful conversations. This work has been supported by NSF PIRE award OISE-1743747. T.T. acknowledges support from the Alfred P. Sloan Foundation and the Ford Foundation.
\end{acknowledgements} 

\bigskip

\bibliography{references}
\end{document}